**Particulate Matter Exposure and Lung Cancer: A Review of two Meta-Analysis Studies.**
S. Stanley Young, CGStat
Warren Kindzierski, University of Alberta

## Abstract (178 words)

The current regulatory paradigm is that PM2.5, over time causes lung cancer. This claim is based on cohort studies and meta-analysis that use cohort studies as their base studies. There is a need to evaluate the reliability of this causal claim. Our idea is to examine the base studies with respect to multiple testing and multiple modeling and to look closer at the meta-analysis using p-value plots. For two meta-analysis we investigated, some extremely small p-values were observed in some of the base studies, which we think are due to a combination of bias and small standard errors. The p-value plot for one meta-analysis indicates no effect. For the other meta-analysis, we note the p-value plot is consistent with a two-component mixture. Small p-values might be real or due to some combination of p-hacking, publication bias, covariate problems, etc. The large p-values could indicate no real effect, or be wrong due to low power, missing covariates, etc. We conclude that the results are ambiguous at best. These meta-analyses do not establish that PM2.5 is causal of lung tumors.

## Introduction

The US Environmental Protection Agency and the World Health Organization claim that PM2.5 is causal of the development of lung cancer deaths. Both support and fund research on air quality and health effects. The standard air components – particulate matter (PM10 and PM2.5), nitrogen dioxide (NO2), and ozone (O3) – are commonly selected as "causes" and all-cause and cause-specific mortalities selected as outcomes.

### Data
Raaschou-Nielsen et al. (2013) did a meta-analysis of 14 cohort studies that examined PM2.5 and lung cancer. Their base cohort studies primarily come from Europe. Hamra et al. (2014) did a systematic review and meta-analysis including 14 cohort studies that examined PM2.5 and lung cancer. Their studies were primarily from the United States.

### Methods: counting
It is important to get some sense of the number of questions under consideration in each cohort study. It is time-consuming and expensive to set up and follow a cohort study. Whereas it is relatively inexpensive to add new measurements and questions to a cohort study. For those reasons, it is typical to have many questions under consideration with a cohort study. Any paper coming from a cohort study might appear focused on one question, but almost always there are many questions at issue.

We count two things, number of papers, and number of foods if there is a food questionnaire component of the study to get a sense of the number of questions. We use Google Scholar to get an estimate of the number of papers that contain the data set name used by the cohort study. That count can be an over count, so for some data set names, we restrict the search of the name to be in the title of the papers. Often a food frequency questionnaire, FFQ, is part of a cohort study. People in the cohort are asked which foods and often how much of those foods are consumed. A FFQ sometimes contains only a few foods, but usually it contains many, 61 to over 200. Clearly, if there are many foods and many health outcomes, we should expect many claims at issue and many resulting papers. So, if a particular paper has only one outcome, in our case lung cancer, and one causative agent, PM2.5, we are seeing the tip of the iceberg.

**Methods: P-value plot**
Briefly, p-values were computed for statistics drawn from base studies into the Raaschou-Nielsen et al. (2013) and the Hamra et al. (2014) meta-analysis. These p-values were ranked from smallest to largest and plotted against the integers to give a p-value plot. See Young and Kindzierski (2018, 2019) for p-value plot formation and other analysis details.

**Methods: Bias and small p-values**
If there is a constant bias, B, in an analysis regardless of the sample size n, then there is a p-value problem as n gets large. This is explained further. The measured (observed) value of a parameter is M and it is equal to the true value T plus any bias, B; M = T + B. So, if we test M with larger and larger samples, and if the real value of T is zero, then T will converge to zero and we will be left with B. As the Standard Error SE gets small, B/SE gets large and we can get an extremely small p-value.

**Methods: Robustness of meta-analysis**
The statistical analysis of a meta-analysis involves combining risk ratios and confidence limits from individual base studies into an estimate of an overall effect with confidence limits. The methods are similar to Fisher's Method of combining of p-values and involve summing terms over the base studies. Any summation method is not robust to outliers so the combining methods will likely declare a strong signal if outliers are present.

**Results**
Counts in Tables 1 and 2 give indication that there are multiple questions at issue in the Raaschou-Nielsen et al. (2013) and Hamra et al. (2014) base cohort studies. The next version will include FFQ information.

P-values were computed from risk ratios and confidence limits and are given in Tables 3 and 4 for Raaschou-Nielsen and Hamra. P-value plots for lung cancer mortality and PM2.5 are given in Figures 1 and 2.

In Figure 1 we see a roughly 45-degree line of points, which is consistent with no effect. In Figure 2 we see the rather common bilinear p-value plot, Young and Kindzierski (2018, 2019). There are p-values below 0.05 and p-values above 0.05 that form an approximate 45-degree line.

The results are ambiguous: the small p-values imply a real effect while the p-values on the 45-degree line imply randomness.

**Discussion**

The two meta-analysis studies appear to be coordinated. Raaschou-Nielsen is the lead author of one study and a coauthor of the other. They are published in consecutive years, 2013 and 2014. Both are funded by government agencies. Despite being a meta-analysis itself, the Raaschou-Nielsen et al. (2013) study is used as a base paper in the Hamra et al. (2014) meta-analysis. The two studies are not independent.

There are two important covariates for lung cancer, smoking and possibly radon. Smoking is an agreed upon cause of lung cancer (Dela Crus et al. 2011) and the cohort studies are expected to take smoking into account. Radon is unusual. The official position on radon is that the higher the level the greater the chance of lung cancer (Dela Crus et al. 2011). This position is based on miners that were occupationally exposed to high levels of radon. Natural radon levels present a much different picture. For natural levels, up to a point, the higher the radon level the lower the lung cancer rate (Obenchain et al. (2019) and references therein).

It is quite possible that many, if not all, of the cohort base studies did not consider radon as a covariate. Radon has an influence on lung cancer and any study not taking it into account is subject to bias. Two of the Hamra base studies mentioned radon, but only Hystad used radon as a covariate. Three of the Raaschou-Nielson base studies mention radon, but none used radon as a covariate.

A causal claim, PM2.5 causes lung cancer, can be disputed on several grounds. The most obvious criticism is dose. Smoking takes many years to induce lung cancer and the daily exposure is ~600 times higher than ambient PM2.5; see Milloy 2016, page 17ff.

Now we list some more technical reasons for disputing the causal claim. i) Any small bias in the base papers used in the meta-analysis might tilt the risk ratio, e.g. radon omitted as a covariate. ii) The papers that were candidates for selection as base papers for the meta-analysis might themselves be a biased set. iii) The data sets used in the base cohort studies have given rise to many papers, from tens to thousands. iv) There could be publication bias; the authors of the base cohort studies likely did not seek to publish negative findings. v) The base papers were checked; they did not correct for multiple testing or multiple modeling.

Note that a base paper with a small *random* p-value would pass a biased risk ratio to the meta-analysis, Young and Kindzierski (2018, 2019). Given our counting of questions/models in other environmental epidemiology papers, Young and Kindzierski (2018, 2019), it is reasonable to think the numbers of questions/models in base papers here are no better than those that have been examined where the median number of questions/models was on the order of 10,000.

Exceedingly small p-values require explanation. The smallest p-value from Table 4 below, Krewski, is $1.03 \times 10^{-5}$, a value so small as to imply certainty. A p-value this small can come about by bias and a small Standard Error. What are some possible sources of bias? High or low temperature can kill the old and the weak. Living in a controlled temperature environment can protect individuals from extreme temperature. There was a heat wave in northern Europe in the year 2003. There were an estimated 70,000 excess deaths as many people did not have air conditioning. Spikes in temperature can kill. High temperatures often occur with high pressure and little wind, and these are conditions that allow poor air quality to develop. PM2.5 can be high while temperature is high.

Multiple factors discussed above – multiple testing and multiple modeling bias, publication bias, bias combined with a small standard error, large negative studies, etc. – present plausible arguments against a claim that PM2.5 is causal of lung cancer. There is no convincing evidence of an effect of PM2.5 on lung cancer mortality in these meta-analysis studies.

**References for Raaschou-Nielson et al. (2013)**

**References for Hamra et al. (2014)**

**Tables and Figures**

Table 1. Cohort/Dataset names are from Raaschou-Nielsen et al. (2013), their Table 1.

| RowID | Cohort/Dataset name | Country | Citations[1] |
|---|---|---|---|
| 1 | EPIC-Umeå | Sweden | 23 |
| 2 | HUBRO | Norway | 31 |
| 3 | SNAC-K | Sweden | 19 |
| 4 | SALT | Sweden | 25 |
| 5 | Sixty | Sweden | 11 |
| 6 | SDPP | Sweden | 37 |
| 7 | DCH | Denmark | 10 |
| 8 | EPIC-MORGEN | Netherlands | 170 |
| 9 | EPIC-PROSPECT | Netherlands | 61 |
| 10 | EPIC-Oxford Cohort | UK | 36,500 |
| 11 | VHM&PP | Austria | 45 |
| 12 | EPIC-Varese | Italy | 51 |
| 13 | EPIC-Turin | Italy | 371 |
| 14 | SIDRIA-Turin | Italy | 14 |
| 15 | SIDRIA-Rome | Italy | 19 |
| 16 | EPIC-San Sebastian | Spain | 17 |
| 17 | EPIC-Athens (air pollution) | Greece | 14 |

[1]Google Scholar was used to determine the approximate number of papers using each data set.

Table 2. Author and cohort names are taken from Hamra et al. (2014).

| RowID | Author | Cohort/Dataset name | Citations[1] |
|---|---|---|---:|
| 1 | Beelen | Netherlands Cohort Study on Diet and Cancer | 1,960 |
| 2 | Cao | China National Hypertension Survey | 98 |
| 3 | Carey | Clinical Practice Research Datalink | 8,380 |
| 4 | Cesaroni | Rome Longitudinal Study | 195 |
| 5 | Hart | Trucking Industry Particle Study | 32 |
| 6 | Hystad | National Enhanced Cancer Surveillance System | 454 |
| 7 | Jerrett | Cancer Prevention Study II | 7,890 |
| 8 | Katanoda | Three-prefecture Cohort Study | 111 |
| 9 | Krewski | Cancer Prevention Study II | 7,890 |
| 10 | Lepeule | Harvard Six Cities Study | 3,530 |
| 11 | Lipsett | California Teachers Study | 3,800 |
| 12 | McDonnell | AHSMOG | 731 |
| 13 | Puett | Nurses' Health Study (in title) | 1,460 |
| 14 | Raaschou[2] | European Study of Cohorts for Air Pollution Effects | 1,940 |

[1]Google Scholar was used to determine the approximate number of papers using each data set.
[2]This study is itself a meta-analysis study.

Table 3. Raaschou-Nielsen et al. (2013) PM2.5 Risk Ratios, Confidence Limits, p-values.

| RowID | Ref | RR | CLlow | CLhigh | SE95 | Z95 | p-value | Rank |
|---|---|---|---|---|---|---|---|---|
| 1 | HUBRO | 0.83 | 0.35 | 2.00 | 0.4209 | -0.4039 | 0.6863 | 8 |
| 2 | SNAC-K | 0.73 | 0.12 | 4.37 | 1.0842 | -0.2490 | 0.8033 | 11 |
| 3 | SALT | 1.24 | 0.23 | 6.76 | 1.6658 | 0.1441 | 0.8854 | 13 |
| 4 | Sixty | 1.56 | 0.41 | 5.98 | 1.4209 | 0.3941 | 0.6935 | 9 |
| 5 | SDPP | 2.01 | 0.40 | 10.01 | 2.4515 | 0.4120 | 0.6803 | 7 |
| 6 | DCH | 0.91 | 0.52 | 1.60 | 0.2755 | -0.3267 | 0.7439 | 10 |
| 7 | EPIC-MORGEN | 0.49 | 0.08 | 3.21 | 0.7985 | -0.6387 | 0.5230 | 5 |
| 8 | EPIC-PROSPECT | 1.09 | 0.17 | 6.99 | 1.7398 | 0.0517 | 0.9587 | 14 |
| 9 | EPIC-Oxford | 0.53 | 0.15 | 1.91 | 0.4490 | -1.0468 | 0.2952 | 2 |
| 10 | VHM&PP | 1.32 | 0.97 | 1.81 | 0.2143 | 1.4933 | 0.1353 | 1 |
| 11 | EPIC-Turin | 1.60 | 0.67 | 3.81 | 0.8010 | 0.7490 | 0.4538 | 3 |
| 12 | SIDRIA-Turin | 1.94 | 0.54 | 7.00 | 1.6480 | 0.5704 | 0.5684 | 6 |
| 13 | SIDRIA-Rome | 1.33 | 0.69 | 2.58 | 0.4821 | 0.6844 | 0.4937 | 4 |
| 14 | EPIC-Athens | 0.90 | 0.34 | 2.40 | 0.5255 | -0.1903 | 0.8491 | 12 |

Table 4. Hamra et al. (2014) PM2.5 Risk Ratios, Confidence Limits, p-values.

| RowID | Author | Year | RR | CLlow | CLhigh | SE | Z | p-value | Rank |
|---|---|---|---|---|---|---|---|---|---|
| 1 | Beelen | 2008 | 0.81 | 0.63 | 1.04 | 0.1046 | -1.8166 | 0.069281 | 5 |
| 2 | Cao | 2011 | 1.24 | 1.12 | 1.37 | 0.0638 | 3.7632 | 0.000168 | 2 |
| 3 | Carey | 2013 | 1.11 | 0.86 | 1.43 | 0.1454 | 0.7565 | 0.449355 | 12 |
| 4 | Cesaroni | 2013 | 1.05 | 1.01 | 1.10 | 0.0230 | 2.1778 | 0.029423 | 3 |
| 5 | Hart | 2011 | 1.18 | 0.95 | 1.48 | 0.1352 | 1.3313 | 0.183083 | 7 |
| 6 | Hystad | 2013 | 1.29 | 0.95 | 1.76 | 0.2066 | 1.4035 | 0.160481 | 6 |
| 7 | Jerrett | 2013 | 1.12 | 0.91 | 1.37 | 0.1173 | 1.0226 | 0.306493 | 10 |
| 8 | Katanoda | 2011 | 1.13 | 0.94 | 1.34 | 0.1020 | 1.2740 | 0.202663 | 8 |
| 9 | Krewski | 2009 | 1.09 | 1.05 | 1.13 | 0.0204 | 4.4100 | 1.03E-05 | 1 |
| 10 | Lepeule | 2012 | 1.37 | 1.07 | 1.75 | 0.1735 | 2.1329 | 0.03293 | 4 |
| 11 | Lipsett | 2011 | 0.95 | 0.70 | 1.28 | 0.1480 | -0.3379 | 0.735415 | 14 |
| 12 | McDonell | 2000 | 1.39 | 0.79 | 2.46 | 0.4260 | 0.9154 | 0.359956 | 11 |
| 13 | Puett | 2014 | 1.06 | 0.90 | 1.24 | 0.0867 | 0.6918 | 0.489085 | 13 |
| 14 | Raaschou-Nielsen[1] | 2013 | 1.39 | 0.91 | 2.13 | 0.3112 | 1.2531 | 0.210164 | 9 |

[1]This study is itself a meta-analysis study.

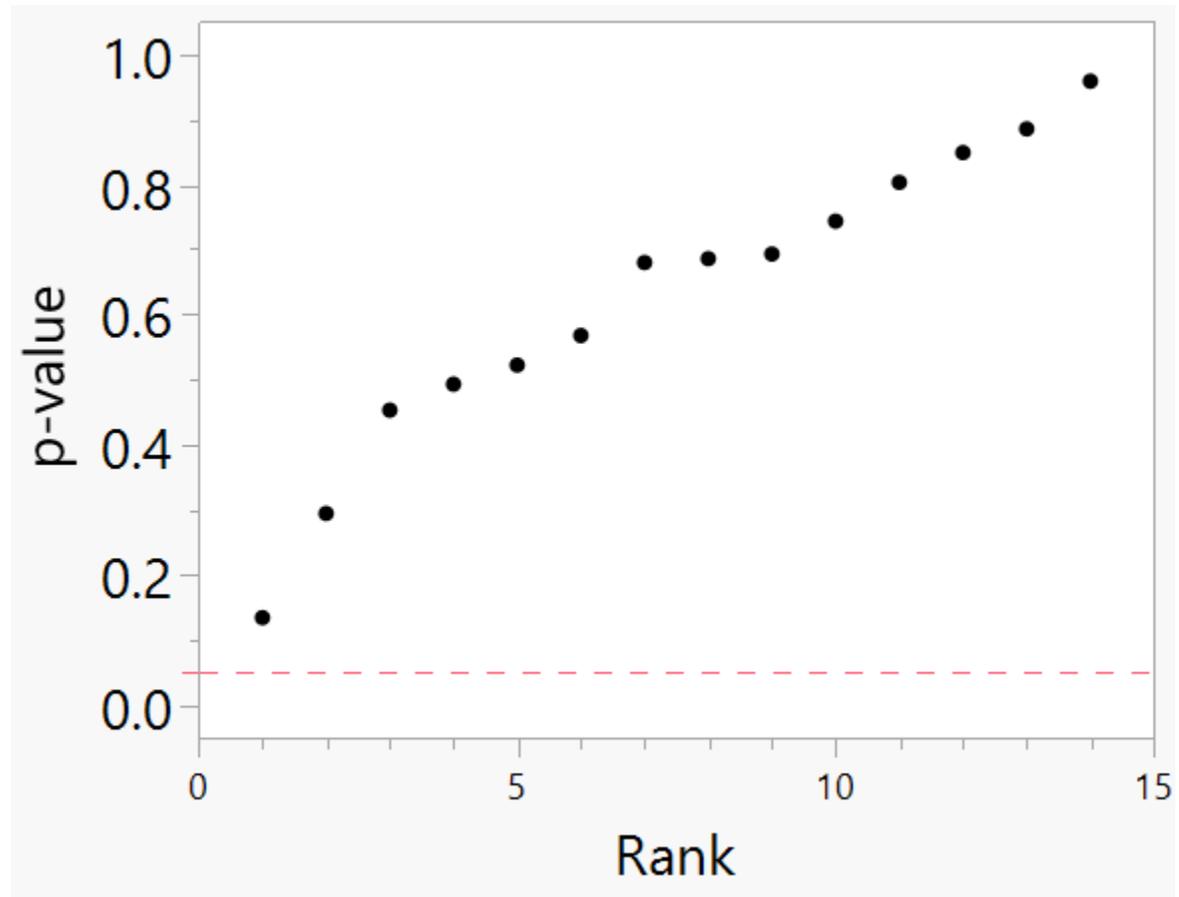

Figure 1. P-value plot of PM2.5 and lung cancer. The data are from Raaschou-Nielsen et al. (2013).

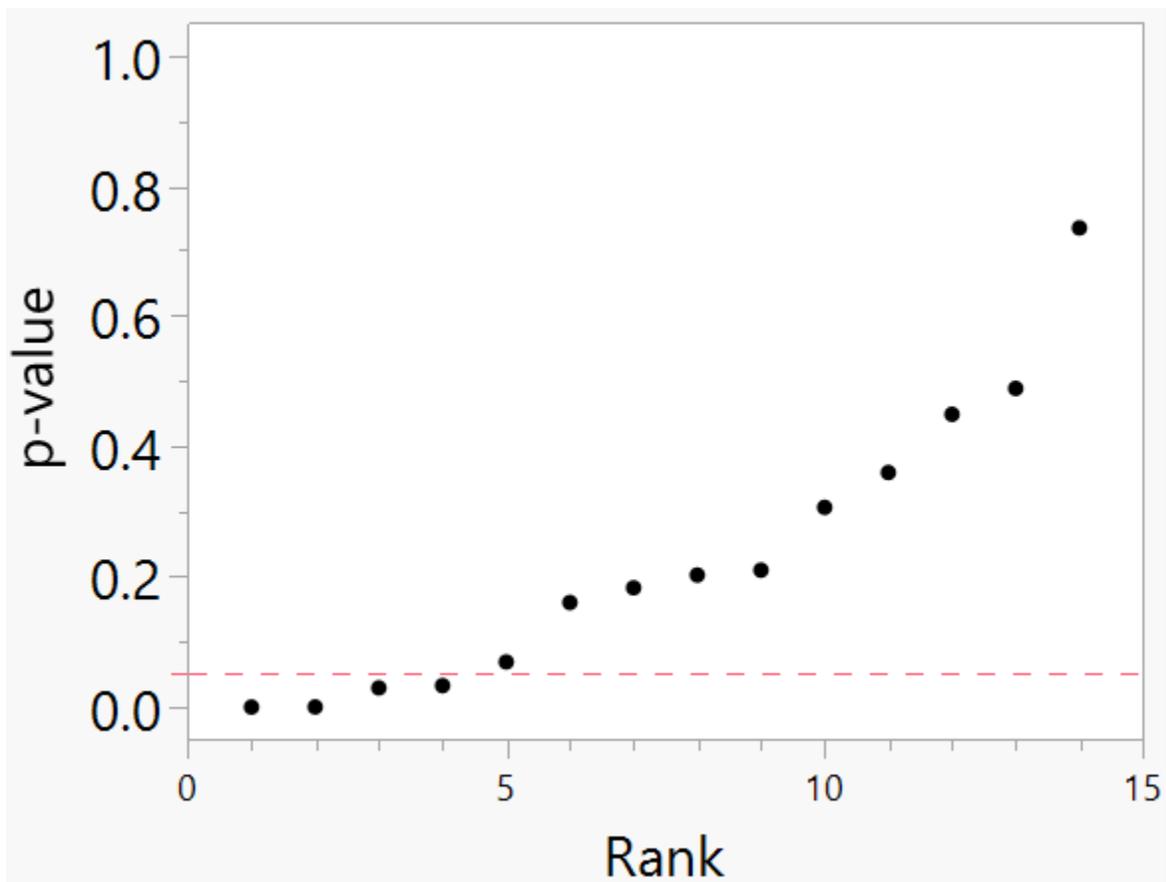

Figure 2. P-value plot of PM2.5 and lung cancer. The data are from Hamra et al. (2014).